\documentstyle {article}
\begin{document}
\textwidth 18.cm
\textheight 23.0cm
\topmargin -0.5in
\baselineskip 16pt
\parskip 18pt
\parindent 30pt
\title{ \large \bf A remark on kinks and time machines}
\author{Ulvi Yurtsever \\
{}~~~~~~~~~~~~ \\
Jet Propulsion Laboratory 169-327\\
California Institute of Technology\\
4800 Oak Grove Drive\\
Pasadena, CA 91109\\
and\\
Theoretical Astrophysics 103-33\\
California Institute of Technology\\
Pasadena, CA 91125}
\date{September, 1994\thanks{Submitted in Physical Review D}}
\pagestyle{empty}
\baselineskip 24pt
\maketitle
\vspace{.2in}
\baselineskip 12pt
\begin{abstract}
\noindent We describe an elementary proof that a manifold with the
topology of the Politzer time machine does not admit a nonsingular,
asymptotically flat Lorentz metric.
\vspace{0.5cm}

\end{abstract}

\newpage
\pagestyle{empty}
\baselineskip 16pt
\parskip 16pt
\pagestyle{plain}
\pagenumbering{arabic}
{}~~~~~~

In a recent paper Chamblin, Gibbons and Steif [1] have used the idea of
gravitational kinks [2--3] to prove that in even spacetime
dimensions there does not exist a
nonsingular, asymptotically flat Lorentz metric on the smooth
Politzer time machine. The $n$-dimensional Politzer time machine is obtained
by (i) starting from the Minkowski spacetime ${\bf R}^n$ with the flat
metric, (ii) by cutting out and duplicating the
two closed unit $(n-1)$--balls centered
at the origin in each of the spacelike
hypersurfaces $\{t=0\}$ and $\{t=1\}$, so that a pair of spherical
``slits" is created, one in each hypersurface, and (iii) by identifying the
upper edge of the first slit with the lower edge of the
second, and the lower edge of the first slit with the upper edge
of the second. The resulting topological space is
not a manifold, as the boundary points of the two slits (the points
that lie on the boundary spheres of the two unit $(n-1)$--balls) do
not have locally Euclidean open neighborhoods. However, these
singularities can be trivially smoothed-out to obtain a manifold with
the same global topology as the original space;
we will call this manifold ``smooth Politzer space"
for emphasis. It has the topology of a $n$--dimensional ``handle", i.e.
$S^{(n-1)} \times S^1$ with a point (which corresponds to infinity)
removed. (A more drastic way to obtain a smooth manifold is to simply
remove the troublesome boundary points of
the two slits; together with the flat metric
on the original space, this yields a well-defined spacetime with
singularities where the boundary points are removed.
It is this space that is usually
referred to as the ``Politzer spacetime" in the literature.)
The question addressed by [1], and also by the present note, is whether a
smooth Lorentz metric, which has
the same qualitative behavior as the flat metric on the original
Politzer space,
exists on the smooth Politzer manifold.
In particular, such a metric needs to be asymptotically flat at the
asymptotically Euclidean ``end" (the point at ``infinity" removed
from $S^{(n-1)} \times S^1$) of the Politzer handle. (For more detailed
background information we refer to [1].)

This question has been answered negatively in [1] by Chamblin, Gibbons
and Steif. The techniques used by these authors, based on the general
notion of gravitational kinks [2--3], have wide scope and great power
in dealing with questions of this kind. However, for the present
specific question about the smooth Politzer space, there exists a simpler
proof of the negative answer that uses only elementary
differential topology; we will now present this proof. Our argument is
based on a single well-known result: the Poincare-Hopf theorem [4],
which states that for a compact, orientable
manifold $M$ and a smooth vector field
$X$ on $M$ with only isolated zeros, the Euler number $\chi (M)$ is
equal to the sum over the zeros of $X$ of the indices of $X$ at those
zeros. One immediate corollary of this result is
that if $\chi (M) \neq 0$, a compact, orientable
$M$ does not admit an everywhere-nonzero
vector field, and hence does not admit a smooth Lorentz metric.
Another corollary is the formula for the Euler number of a connected
sum: If $M_{1}$ and $M_{2}$ are compact,
orientable even ($2m$--)dimensional manifolds,
then $\chi(M_{1} \# M_{2}) = \chi(M_{1}) + \chi (M_{2}) -2 $. To
derive this from the Poincare-Hopf, consider vector fields $X_{1}$
and $X_{2}$ with isolated zeros on $M_{1}$ and $M_{2}$, respectively, so
that $X_{i}$ has a simple zero at $p_{i} \in M_{i}$. Since the antipodal
map on the odd-sphere $S^{(2m-1)}$ is homotopic to identity ([4]), we can,
without changing the indices at the zeros,
flip the signs of the $X_{i}$ if necessary and arrange that
$X_{1}$ is inward-pointing near $p_{1}$ and $X_{2}$ is outward-pointing
near $p_{2}$. Now we perform the connected sum by
joining $M_{1}$ and $M_{2}$ in a neighborhood of $p_{i}
\in M_{i}$. Then we can smoothly extend the vector fields $X_{i}$ to a vector
field $X$ on $M_{1} \# M_{2}$ such that $X$
has precisely the same set of zeros and
indices as the $X_{i}$ except for the simple zeros at $p_{1}$
and $p_{2}$: $X$ is smooth and
nonzero near $p_{1} \equiv p_{2}$ by construction. The formula now follows
by simple counting.

Now assume that the smooth Politzer space $S^1 \times S^{(n-1)}
\; \backslash \; p_{\infty}$ admits a smooth Lorentz metric which is
asymptoptically flat near $p_{\infty}$. Let $T^{n}$ denote the
$n$--torus with the standard flat Lorentz metric on it. Since the
Lorentz metric on the Politzer manifold is asymptotically flat at
$p_{\infty}$, we can combine the two metrics smoothly to obtain a
Lorentz metric on the connected sum $(S^1 \times S^{(n-1)}) \# T^n$.
But the Euler number of $S^1 \times S^{(n-1)}$ and of $T^n$ are both
zero. If the dimension $n$ is even, by the formula above $\chi [ (S^1
\times S^{(n-1)}) \# T^n ] = -2$, accordingly, it is impossibe for
$(S^1 \times S^{(n-1)}) \# T^n$ to admit a smooth Lorentz metric. This
contradiction proves the result we seek in even dimensions $n$.

More generally, let us call a
$n$--manifold $M$ ``asymptotically Euclidean"
if compact subsets $K \subset M$ and $B \subset {\bf R}^n$ exist such that
$M \backslash K$ is diffeomorphic to ${\bf R}^n \backslash B$. Call a Lorentz
metric on $M$ asymptotically flat if on
$M \backslash K$ it is asymptotic to the Minkowski
metric. For an asymptotically Euclidean $M$, let
${M}^{\infty}$ denote the one-point compactification of $M$ obtained by
smoothly adjoining to $M$ the point at infinity
of $M\backslash K$. Thus, if $M$ is
${\bf R}^n$, then $M^{\infty}$ is $S^n$; for the smooth
Politzer space $M^{\infty}$
is $S^1 \times S^{(n-1)}$. The argument above proves the following
result:

\noindent Let $M$ be an orientable, even-dimensional asymptotically Euclidean
manifold. If $\chi (M^{\infty}) \neq 2$, then $M$ does not admit a smooth,
asymptotically flat Lorentz metric.

\newpage

\noindent{\bf REFERENCES}

\noindent{\bf 1.} A. Chamblin, G. W. Gibbons and A. R. Steif,
Phys. Rev. D {\bf 50}, R2353 (1994).

\noindent{\bf 2.} G. W. Gibbons and S. W. Hawking, Phys. Rev. Lett.
{\bf 69}, 1719 (1992).

\noindent{\bf 3.} G. W. Gibbons and S. W. Hawking,
Commun. Math. Phys. {\bf 148}, 345 (1992).

\noindent{\bf 4.} Hirsch, M. W.: Differential Topology. Springer Verlag,
Berlin Heidelberg New York 1976.

\end{document}